\documentclass[12pt, a4paper]{article}
\usepackage[utf8]{inputenc}
\pdfoutput=1

\usepackage[top=1in, bottom=1in, left=0.75in, right=0.75in]{geometry}
\usepackage[T1]{fontenc}
\usepackage{lmodern}
	\DeclareFontFamily{OMX}{lmex}{}
	\DeclareFontShape{OMX}{lmex}{m}{n}{<-> lmex10}{}	
\usepackage{abstract}
\usepackage{amsmath, amssymb, amsfonts, mathtools}	
	\numberwithin{equation}{section}	
\usepackage{graphicx}       
	\graphicspath{ {./figures/} }
\usepackage[caption=false]{subfig} 
\usepackage{caption}   
\usepackage[dvipsnames]{xcolor}         
\usepackage{physics}
\usepackage{tensor}
\usepackage{mathrsfs}
\usepackage{bm}
\usepackage{enumitem}
\usepackage{cite}

\usepackage{yhmath} 	

\usepackage[bottom]{footmisc}	

\usepackage[subfigure]{tocloft}

\usepackage{hyperref}		
\definecolor{myRed}{rgb}{0.545,0,0}
\definecolor{myDarkBlue}{rgb}{0,0,0.5}
\hypersetup{colorlinks=true, linktocpage, linkcolor=blue, citecolor={myRed}, urlcolor=blue}  

\def \be {\begin{equation}}
\def \ee {\end{equation}}
\def \nn {\nonumber}
\def \del {\partial}

\DeclarePairedDelimiterX\inp[2]{\langle}{\rangle}{#1\,\delimsize\vert\,\mathopen{}#2}	   

\begin{document}

\begin{titlepage}
\vspace*{-10mm}   
\baselineskip 10pt      
\baselineskip 20pt   

\begin{center}
\noindent
{\LARGE\bf
Hawking Radiation Under \\
Generalized Uncertainty Principle\par}
\vskip10mm
\baselineskip 20pt

\renewcommand{\thefootnote}{\fnsymbol{footnote}}

{\large
Tin-Long~Chau$^a$
\footnote[1]{\texttt{\url{ronnychau031@gmail.com}}},
Pei-Ming~Ho$^{a, b}$
\footnote[2]{\texttt{\url{pmho@phys.ntu.edu.tw}}},
Hikaru~Kawai$^{a, b}$
\footnote[4]{\texttt{\url{hikarukawai@phys.ntu.edu.tw}}},\\
Wei-Hsiang~Shao$^a$
\footnote[5]{\texttt{\url{whsshao@gmail.com}}}, and 
Cheng-Tsung~Wang$^a$
\footnote[6]{\texttt{\url{ctgigglewang@gmail.com}}}
}

\renewcommand{\thefootnote}{\arabic{footnote}}

\vskip5mm

{\normalsize \it  
$^a$Department of Physics and Center for Theoretical Physics,\\
National Taiwan University, Taipei 10617, Taiwan
\\
$^b$Physics Division, National Center for Theoretical Sciences, Taipei 10617, Taiwan
}  

\vskip 15mm
\begin{abstract}
\vspace{-3mm}
\normalsize

The generalized uncertainty relation is expected to be an essential element 
in a theory of quantum gravity. 
In this work, 
we examine its effect on the Hawking radiation 
of a Schwarzschild black hole formed from collapse 
by incorporating a minimal uncertainty length scale 
into the radial coordinate of the background.
This is implemented in both the ingoing Vaidya coordinates 
and a family of freely falling coordinates. 
We find that, 
regardless of the choice of the coordinate system, 
Hawking radiation is turned off at around the scrambling time. 
Interestingly, 
this phenomenon occurs while the Hawking temperature remains largely unmodified.

\end{abstract}
\end{center}

\end{titlepage}

\newcommand\afterTocSpace{\bigskip\medskip}
\newcommand\afterTocRuleSpace{\bigskip\bigskip}

\hrule
\tableofcontents
\afterTocSpace
\hrule
\afterTocRuleSpace

\section{Introduction}

Hawking's seminal discovery~\cite{Hawking:1974rv, Hawking:1975vcx} 
of the evaporation of black holes 
has given rise to the information loss problem~\cite{Hawking:1976ra, Mathur:2009hf}, 
the resolution of which is expected to offer new insights into quantum gravity. 
However, 
the conventional description~\cite{Hawking:1975vcx} of Hawking radiation 
within the framework of effective field theory 
involves trans-Planckian modes, 
thus leaving some room of doubt 
regarding the validity of its predictions~\cite{tHooft:1984kcu, Jacobson:1991gr}.
In particular, 
recent investigations~\cite{Ho:2020cbf, Ho:2020cvn, Ho:2021sbi, Ho:2022gpg} 
have shown that the inclusion of 
higher-derivative (non-renormalizable) couplings 
between the radiation field and background curvature 
leads to a breakdown of the effective field theory 
beyond the \emph{scrambling time}\,\footnote{
For a black hole with Schwarzschild radius $a$, 
the scrambling time is defined to be 
$a \log (a / \ell_p)$~\cite{Sekino:2008he}, 
where $\ell_p$ is the Planck length.
},
owing to the genuinely high (Lorentz-invariant) 
energy scales of the processes involved.  
With that being said, 
there is still a widespread belief that
Hawking radiation, as predicted by the low-energy effective theory, 
remains robust against physics in the UV regime.
This idea has garnered additional support over the years 
through examining a variety of UV models~\cite{Unruh:1994je, Brout:1995wp, Hambli:1995pp, Corley:1996ar, Corley:1997ef, Corley:1997pr, Jacobson:1999ay, Unruh:2004zk, Barcelo:2005fc, Agullo:2009wt, Kajuri:2018myh}. 
Nevertheless, 
there are also studies that entertain the possibility 
of a different outcome~\cite{Jacobson:1993hn, Helfer:2003va, Barcelo:2008qe, Akhmedov:2015xwa, Ho:2022gpg, Akhmedov:2023gqf}.

A common theme across various approaches to quantum gravity 
is the emergence of a fundamental minimum length scale 
$\ell$~\cite{Garay:1994en, Ali:2009zq}, 
such as the string length or the Planck length, 
below which the usual notion of spacetime breaks down.
Since the precise quantum-gravitational effects at this scale remain unknown, 
in order to address the implications of a minimum length,
a pragmatic approach often employed is to introduce it through 
the \emph{Generalized Uncertainty Principle} (GUP):\,\footnote{
We will work with natural units $\hbar = c = 1$ throughout this paper.
}
\be
\Delta x \Delta p 
\geq 
\frac{1}{2} \left[ 1+\ell^2 ( \Delta p )^2 \right] \, ,
\label{GUP}
\ee
where $\Delta x$ and $\Delta p$ represent the uncertainties 
in position and momentum, respectively.
This relation is motivated by string theory~\cite{Amati:1987wq, Gross:1987kza, Gross:1987ar, Amati:1988tn, Fabbrichesi:1989ps, Konishi:1989wk, Guida:1990st} 
as well as general considerations of quantum gravity~\cite{Maggiore:1993rv, Scardigli:1999jh, Adler:1999bu, Capozziello:1999wx, Scardigli:2003kr},
and it results in a minimal uncertainty $(\Delta x)_{\text{min}} = \ell$ in position.

In this work, 
we aim to study the impact of the GUP~\eqref{GUP} on the Hawking radiation 
of a large dynamical black hole ($a \gg \ell$) with spherical symmetry. 
The minimum length $\ell$ is introduced into the position $x$ 
representing the radial coordinate of this background, 
which we implement by deforming the commutator 
in the Heisenberg algebra as~\cite{Kempf:1994su, Kempf:1996ss, Kempf:1996nk}\,\footnote{
In general, 
the commutation relation can take the form 
$\left[ \hat{x},\hat{p} \right] = i \left( 1 + \alpha \, \hat{x}^2 + \beta \, \hat{p}^2 \right)$ 
for some constants $\alpha$ and $\beta$. 
We set $\alpha = 0$ and $\beta = \ell^2$ here for simplicity,
which leads to the uncertainty relation 
$\Delta x \Delta p \geq \left[ 1 + \ell^2 (\Delta p)^2 + \ell^2 \langle p \rangle^2 \right] / 2$,
thus implying eq.~\eqref{GUP}.
For further information on the operator algebra, Hilbert space, 
and other aspects of quantum theory related to the GUP, 
see Refs.~\cite{Kempf:1994su, Kempf:1996ss, Kempf:1996nk, Detournay:2002fq, Segreto:2022clx}.
}
\be 
\label{GUP-comm}
\left[ \hat{x},\hat{p} \right] = i \left( 1+\ell^2 \hat{p}^2 \right) \, .
\ee 
This modification can be realized in momentum space 
via the representation~\cite{Kempf:1994su} 
\be 
\label{hat_x_hat_p}
\hat{x} \equiv i \left( 1 + \ell^2 p^2 \right) \del_p \, , \qquad \hat{p} = p \, .
\ee 
Within this framework, 
the corrections induced by the GUP are encoded 
in the wave equation governing the radiation field, 
leading to significant alterations of the behavior of Hawking modes 
that probe the black hole background with trans-Planckian momenta. 

This GUP framework has been previously applied~\cite{Brout:1998ei} 
to investigate Hawking radiation 
in the Eddington-Finkelstein coordinates, 
revealing intriguing effects on wave packet propagation. 
Furthermore, 
there are extensive studies in the literature 
on corrections to the Unruh temperature~\cite{Scardigli:2018jlm}, 
Hawking temperature,
and other thermodynamic properties of black holes 
within the context of GUP~\cite{Amelino-Camelia:2005zpp, Nozari:2008gp, Majumder:2012rtc, Chen:2013ssa, Miao:2014jea, Wang:2014cza, Chen:2014xgj, Bargueno:2015tea, Mu:2015qta, Sakalli:2016mnk, Gecim:2017zid, Lambiase:2017adh, Kanazawa:2019llj, Buoninfante:2019fwr, IbungochoubaSingh:2019nzh, Kanzi:2021jrl, Anacleto:2021nhm, Anacleto:2022lnt, Anacleto:2022sim, Ong:2023jkp, Anacleto:2023ali}. 
In contrast, 
in this work we shall focus on the 
time-dependent magnitude of Hawking radiation.
Notably, 
we discover that 
Hawking radiation maintains a thermal spectrum
\be 
\ev{\mathcal{N}_{\omega}(t)}
\approx 
\frac{1}{e^{\omega / T_H} - 1} \, \mathcal{A}(t)
\ee 
characterized by essentially the same Hawking temperature $T_H = 1 / 4 \pi a$,
but with an amplitude $\mathcal{A}(t)$ 
that diminishes significantly after the scrambling time:\,\footnote{
A similar phenomenon was found in Ref.~\cite{Ievlev:2023inj},
in which a moving point charge emits thermal Larmor radiation 
at a finite temperature but with an exceedingly small amplitude 
if its final speed is sufficiently slow.
}
\be
\mathcal{A}(t) \to 0 
\quad \text{for} \quad
t \gtrsim 2a \log(a / \ell) \, ,
\label{t=scr}
\ee
where $t$ denotes the duration in Schwarzschild time, 
starting from the moment the collapsing matter is at a distance 
of $\mathcal{O}(a)$ from the black hole horizon.

The suppression of Hawking radiation at late times 
can be attributed to the substantial change 
in the behavior of $\Delta x$ in the domain of 
large $\Delta p \gtrsim \ell^{-1}$ caused by the GUP~\eqref{GUP}. 
In the low-energy effective theory 
which assumes locality and 
a reciprocal relation between the uncertainties 
$\Delta x$ and $\Delta p$, 
an outgoing wave packet 
with purely positive frequency at a large distance 
can be traced back in time to the near-horizon region, 
where the central momentum of the wave packet 
experiences an exponential blueshift.
This causes the packet to be densely compressed 
close to the horizon in the past 
with $\Delta x \lesssim \ell$.
Consequently, 
the wave packet has a broad spectrum $\Delta p$ in momentum, 
consisting of both positive and negative-$p$ components. 
This mixture is the key ingredient for the production of Hawking radiation. 

In the case of the GUP~\eqref{GUP}, however, 
a remarkable observation is that in fact 
$\Delta x \gg a$ due to the immensely large spread 
in momentum $\Delta p \gg a / \ell^2$ 
when the wave packet is traced back 
to a distance of $\mathcal{O}(\ell)$ from the horizon.
Not only does this imply that the wave packet 
is no longer confined within the near-horizon region, 
it also suggests that the wave packet might even 
largely surpass the size of the black hole itself, 
to the extent where its evolution is insensitive to the black hole spacetime.
As a result, 
modes with trans-Planckian momenta 
do not contribute to Hawking radiation, 
leading to the termination of radiation after the scrambling time. 

This paper is organized as follows. 
In Sec.~\ref{sec:vaidya}, 
we incorporate the GUP into 
the free field theory in the ingoing Vaidya spacetime. 
This is achieved by formulating the theory in momentum space 
and employing the representation~\eqref{hat_x_hat_p}.
We find that Hawking radiation is turned off 
after the scrambling time, as defined in eq.~\eqref{t=scr}.
In Sec.~\ref{sec:ff},
we extend our analysis by introducing the GUP 
into a family of freely falling coordinates. 
Although the implementation of the GUP~\eqref{GUP} 
depends on the choice of the reference frame, 
we arrive at the same conclusion 
regarding the late-time suppression of Hawking radiation.
Finally, 
we wrap up the discussion with a summary 
and concluding remarks in Sec.~\ref{sec:conclusion}.

\section{GUP in Vaidya Coordinates}
\label{sec:vaidya}

In this section,
we adopt the approach outlined in Ref.~\cite{Brout:1998ei}
to study how Hawking radiation is affected by 
the GUP defined with respect to the ingoing Vaidya coordinates. 
Our findings indicate that the incorporation of GUP 
leads to the termination of Hawking radiation 
beyond the scrambling time.

Let us begin our discussion by establishing the setup of the problem.
Recall that a four-dimensional spherically-symmetric black hole 
with Schwarzschild radius $a$ can be described by the metric\,\footnote{
For a black hole of mass $M$, 
$a = 2 G_N M$, 
with $G_N = \ell_p^2$ being the Newton constant in natural units.
}
\be 
ds^2 = 
- \left(1 - \frac{a}{r} \right) dt^2 
+ \left(1 - \frac{a}{r} \right)^{-1} dr^2 
+ r^2 d\Omega^2 \, ,
\label{Schwarzschild}
\ee 
where $d\Omega^2$ stands for the differential solid angle on a unit 2-sphere.
Through a coordinate transformation, 
this metric can be expressed as
\be
ds^2 
= - \left(1 - \frac{a}{r} \right) dv^2 + 2 dvdr + r^2 d\Omega^2 
\ee
in terms of the Eddington-Finkelstein coordinates, 
where $v(t, r) \equiv t + r_*$ 
and the tortoise coordinate $r_*$ is defined by 
\be
r_{\ast} \equiv r + a\log\left( \frac{r}{a} - 1 \right) \, .
\ee
Outgoing waves travelling at the speed of light 
are purely functions of the Eddington retarded time 
$u(t, r) \equiv t - r_*$.  

We consider a simple model where the black hole is formed 
from a thin shell of spherical matter collapsing at the speed of light. 
It is then natural to describe the geometry 
using the ingoing Vaidya metric
\be 
\label{ingoing-Vaidya-metric}
ds^2 = -\left( 1 - \Theta(v) \, \frac{a}{r} \right) dv^2 + 2 dv dr \, ,
\ee  
where $\Theta(v)$ represents the step function,
and for simplicity, 
we have omitted the angular part of the metric.
Without loss of generality,
we have assumed that the trajectory of 
the infalling null shell is given by $v = 0$. 
Outside the shell ($v > 0$), 
the metric corresponds to the Schwarzschild metric, 
with the event horizon located at $r = a$.
In the region inside the shell ($v < 0$), 
the geometry is that of Minkowski space. 

In the near-horizon region of the Vaidya background where
\be
x \equiv r - a \ll a \, ,
\label{x<<a}
\ee 
the action for a massless real scalar field in the low-energy effective theory is given by 
\be 
\label{eq:action0}
S_0 = 
- \frac{1}{2} \int dv \int dx 
\left[ \del_x \phi(v, x) \right]
\biggl\{ 
2 \del_v + \left[ 1 + \Theta(v) \left( \frac{x}{a} - 1 \right) \right] \del_x 
\biggr\} 
\phi(v, x) \, .
\ee 
Via the Fourier transform
\be
\phi(v, x) = 
\int_{-\infty}^{\infty} \frac{dp}{\sqrt{2\pi}} \, \tilde{\phi}(v, p) \, e^{i p x} \, ,
\label{FT}
\ee
the action~\eqref{eq:action0} is equivalent to
\begin{align}
S_0 &= 
\int dv \left\langle \tilde{\phi}(v, p)
\left| \,
p \left[
i \del_v - \frac{p}{2} 
- \Theta(v) 
\left( \frac{\hat{x}}{2a} \, p - \frac{p}{2} \right) 
\right] 
\right|
\tilde{\phi}(v, p) \right\rangle_0 \, ,
\label{S0}
\end{align}
where $\hat{x} \equiv i \del_p$ and the definition of $\inp{A}{B}_0$ is
\be
\inp{A}{B}_0
\equiv
\int_{-\infty}^{\infty} dp \, A^*(p) B(p) \, .
\label{inp0}
\ee
Note that the relativistic inner product conserved under time evolution is
$\langle \tilde{\phi}_1 | \, p \, | \tilde{\phi}_2\rangle_0$.

To accommodate the GUP in this setup, 
we shall utilize the momentum-space representation 
(see for example Refs.~\cite{Damour:1976jd, Brout:1995wp}) and, 
without loss of generality,
realize the modified commutator algebra~\eqref{GUP-comm} 
by adopting eq.~\eqref{hat_x_hat_p}. 
It then follows that eq.~\eqref{inp0} should also be modified as
\be 
\inp*{A}{B} \equiv 
\int_{-\infty}^{\infty} \frac{dp}{1 + \ell^2 p^2} \, 
A^* (p) B(p) \, ,
\label{eq:inp}
\ee 
where a measure factor $1/(1 + \ell^2 p^2)$ has been introduced 
to ensure the Hermiticity of $\hat{x}$~\eqref{hat_x_hat_p} 
on the domain of functions that decay at $\pm \infty$~\cite{Kempf:1994su}.

Following Ref.~\cite{Brout:1998ei},
the GUP modification of the action~\eqref{S0} 
for a massless real scalar is obtained by 
employing the representation~\eqref{hat_x_hat_p} 
and also replacing $\inp{A}{B}_0$ in the action 
with $\inp{A}{B}$ as defined in eq.~\eqref{eq:inp}, 
incorporating the appropriate measure factor. 
This results in 
\begin{align}
S_{\mathrm{GUP}} 
&=
\int dv \int_{-\infty}^{\infty} \frac{dp}{1 + \ell^2 p^2} \,
\tilde{\phi}^{\ast}(v, p) \,
p 
\left[
i \del_v - \frac{p}{2} 
- \Theta(v) 
\left( \frac{\hat{x}}{2a} \, p - \frac{p}{2} \right) 
\right] 
\tilde{\phi}(v, p)
\nn \\
&= 
2 \int dv \int_0^{\infty} \frac{dp}{1 + \ell^2 p^2} \, 
\tilde{\phi}^{\ast}(v, p) \,
p 
\left\{ 
i \del_v - \frac{p}{2} 
- \Theta(v) 
\left[ \frac{i (1 + \ell^2 p^2)}{2a} \, \del_p \, p - \frac{p}{2} \right]
\right\} 
\tilde{\phi}(v, p) \, ,
\label{eq:action}
\end{align}
where in the second line 
we performed a partial integration 
and then imposed the reality condition
\be
\tilde{\phi}^{\ast}(v,p) = \tilde{\phi}(v,-p) 
\ee
on the real scalar field.
Variation of the action~\eqref{eq:action} 
leads to the field equation 
\be 
\label{eq:waveeq}
\left\{
i \del_v - \frac{p}{2} 
- \Theta(v) 
\left[ \frac{i \left( 1 + \ell^2 p^2 \right)}{2a} \, \del_p \, p - \frac{p}{2} \right] 
\right\} 
\tilde{\phi}(v, p) = 0 \, ,
\ee 
which demands the continuity of $\tilde{\phi}(v, p)$ 
across the null shell at $v = 0$. 
According to the field equation,
the $v$-derivative of the relativistic inner product 
of any two solutions $\tilde{\phi}_1$ and $\tilde{\phi}_2$ 
is
\begin{align}
\del_v \langle \tilde{\phi}_1(v) | \, p \, | \tilde{\phi}_2(v) \rangle
&= 
\frac{\Theta(v)}{2a} \, 
\Bigl[ 
p^2 \, \tilde{\phi}_1^{*}(v, p) \, \tilde{\phi}_2(v, p) 
\Bigr]\Big\vert^{\infty}_{p \, = \, -\infty} \, .
\end{align}
Hence, 
the inner product is conserved under time evolution
for solutions satisfying the boundary condition
\be
\lim_{p \rightarrow \pm \infty}
p \, \tilde{\phi}(v, p) = 0 \, .
\label{BC-p}
\ee

Now let us consider the general solution to the field equation~\eqref{eq:waveeq}
inside and outside the shell, respectively.
In the flat spacetime inside the collapsing shell ($v < 0$), 
the equation is simply
\be
\left(i \del_v - \frac{p}{2} \right) \tilde{\phi}(v, p) = 0 \, ,
\ee
and its general solution can be written as
\be 
\label{eq:decomp_in0}
\tilde{\phi}(v < 0, p) = 
e^{-i p v / 2} \sqrt{\frac{1 + \ell^2 p^2}{2|p|}} 
\left[ a_p \, \Theta(p) + a_{-p}^{\dagger} \, \Theta(-p) \right] \, .
\ee  
Upon quantization, 
where the field $\tilde{\phi}$ and its conjugate momentum 
\be 
\widetilde{\Pi}(v, p) = \frac{2 i p}{1 + \ell^2 p^2} \, \tilde{\phi}^{\ast}(v, p)
\ee 
are promoted to operators satisfying 
the equal-time commutation relation 
$\comm*{\tilde{\phi}(v, p)}{\widetilde{\Pi}(v, p')} = i \, \delta(p - p')$, 
the creation and annihilation operators obey
\be 
\comm*{a_p}{a_{p'}^{\dagger}} = \delta(p - p') \, , \qquad \comm{a_p}{a_{p'}} = 0 \, , \qquad \comm*{a_p^{\dagger}}{a_{p'}^{\dagger}} = 0 \, .
\ee 
The quantum state of the field $\phi$ inside the shell
is assumed to be the Minkowski vacuum $\ket{0}$ 
defined by 
\be
\ket{0} \; \ni \;
a_p \ket{0} = 0 \quad \forall \, p > 0 \, .
\label{vacuum}
\ee

In the Schwarzschild spacetime outside the shell ($v > 0$), 
the field equation \eqref{eq:waveeq} takes the form
\be
\left[ \del_v - \frac{(1 + \ell^2 p^2)}{2a} \, \del_p \, p \right]
\tilde{\phi}(v, p) = 0 \, ,
\label{outgoing-wave-eq}
\ee
and its general solution is given by
\begin{align}
\tilde{\phi}(v > 0, p) = p^{-1} \, \Psi \bigl( v + 2a F(p) \bigr)
\label{outgoing-genera-sol}
\end{align}
for an arbitrary function $\Psi$,
where
\be 
F(p) \equiv \log \left( \frac{ a \abs{p}}{\sqrt{1 + \ell^2 p^2}} \right) \, .
\label{F-def}
\ee 
A special case is the single-frequency solution 
describing a Hawking mode:
\be 
\label{eq:wavesoln}
\tilde{\phi}_{\omega}(v, p) = 
\mathcal{N}_{\omega} \, 
\frac{e^{-i \omega v}}{p - i 0^+} 
\left[ \frac{ a \left( p - i 0^+ \right)}{\sqrt{1 + \ell^2 p^2}} \right]^{-2 i a \omega} \, ,
\ee 
where $\omega$ is the Killing frequency 
with respect to the advanced time coordinate $v$, 
and $\mathcal{N}_{\omega}$ is an overall constant 
that will be fixed shortly.
The prescription $p \to p - i 0^+$ for analytic continuation is introduced so that
$\tilde{\phi}_{\omega} (v, p)$ corresponds to an outgoing wave outside the horizon~\cite{Damour:1976jd} (see also Ref.~\cite{Akhmedov:2023gqf}).

Given the characteristic trajectory 
\be 
\label{eq:chr_traj}
v + 2 a F(p) = \text{constant}
\ee
inferred from the general solution~\eqref{outgoing-genera-sol}, 
it is evident that at sufficiently large values of $v$, 
the prevailing momentum is very sub-Planckian, 
i.e., $\abs{p} \ll \ell^{-1}$, 
and it is thus reasonable to treat $\ell p \to 0$ in this regime. 
As a result, 
a typical Hawking mode~\eqref{eq:wavesoln} 
with frequency $\omega \sim 1 /a \ll \ell^{-1}$ 
essentially evolves in accordance with
the low-energy effective theory at large $v$, 
and is expected to coincide with the outgoing plane wave 
$e^{- i \omega u} / \sqrt{2 \omega} \approx e^{-i \omega v} \left( x / a \right)^{2 i a \omega} / \sqrt{2 \omega}$. 
It is then natural to choose the normalization constant $\mathcal{N}_{\omega}$ such that~\cite{Brout:1998ei} 
\be 
\phi_{\omega}(v, x)
\approx
\mathcal{N}_{\omega} \, 
\int_{-\infty}^{\infty} \frac{dp}{\sqrt{2\pi}} \,
e^{i p x} \, \frac{e^{-i \omega v}}{p - i 0^+} 
\left[a(p - i 0^+) \right]^{-2 i a \omega}
= 
\frac{e^{- i \omega v}}{\sqrt{2\omega}} \left( \frac{x}{a} \right)^{2 i a \omega} \, ,
\label{phi-asymp}
\ee 
where $\phi_{\omega}(v, x)$ is related to $\tilde{\phi}_{\omega}(v, p)$ 
via the Fourier transform~\eqref{FT}.
Since $\mathcal{N}_{\omega}$ is independent of both $p$ and $v$, 
it can be uniquely fixed by this equation to be
\be 
\mathcal{N}_{\omega} 
\approx
a 
\sqrt{\frac{\omega}{\pi}} \, 
e^{\pi a \omega} \, 
\Gamma(2i a \omega)
\quad \mbox{for} \quad 
\omega \ll \ell^{-1} \, ,
\label{N-def}
\ee 
which leads to
\be
\left|\mathcal{N}_{\omega}\right|^2
\approx \frac{a}{1 - e^{- 4\pi a\omega}} \, .
\label{N2}
\ee

In order to capture the time dependence of Hawking radiation, 
we consider a \emph{localized} Hawking particle with a wave function 
represented by an outgoing wave packet 
constructed as a superposition of 
the monochromatic solutions~\eqref{eq:wavesoln}:
\be 
\widetilde\Psi_{(\omega_0, u_0)} (v, p) = 
\int_0^{\infty} d \omega \, f_{\omega_0}(\omega) \, 
e^{i \omega u_0} \, \tilde{\phi}_{\omega}(v, p) \, .
\label{vaidya_packet}
\ee 
This packet has central frequency $\omega_0$ 
and is centered around a given Eddington retarded time $u = u_0$. 
The profile function $f_{\omega_0}(\omega)$ is assumed 
to possess a narrow width $\Delta \omega$ such that 
$|f_{\omega_0}(\omega)|$ is negligible when 
$|\omega - \omega_0| \gg \Delta \omega$\,\footnote{
Frequently used profiles include the Gaussian profile 
and the step function profile $f_{\omega_0}(\omega) = \left[ \Theta(\omega - \omega_0 + \Delta \omega / 2) - \Theta(\omega - \omega_0 - \Delta \omega / 2) \right] / \sqrt{\Delta \omega}$ originally introduced by Hawking~\cite{Hawking:1975vcx}.
}.
Moreover, 
$f_{\omega_0}(\omega)$ is suitably normalized so that 
\be 
\label{f2=1}
\int_0^{\infty} d \omega \, \abs{f_{\omega_0}(\omega)}^2 = 1 \, .
\ee 

To better understand the wave packet construction~\eqref{vaidya_packet}, 
recall that at large values of $v$,
the effects of the GUP can be neglected 
due to the dominant momentum being small ($\ell p \ll 1$).
Consequently, 
the mode solutions $\phi_{\omega}$ in the superposition 
can be approximated by outgoing plane waves 
as demonstrated in eq.~\eqref{phi-asymp}, 
and the position-space representation 
of the wave packet~\eqref{vaidya_packet} 
can be simplified as
\be
\int_{-\infty}^{\infty} \frac{dp}{\sqrt{2 \pi}} \, e^{i p x} \, \widetilde\Psi_{(\omega_0, u_0)} (v \gg u_0 \, , p)
\approx
\frac{1}{\sqrt{2\omega_0}}
\int_0^{\infty} d \omega \, f_{\omega_0}(\omega) \, 
e^{- i \omega (u - u_0)} \equiv \Psi_{\omega_0}(u - u_0) \, .
\label{Psi=Phi-1}
\ee
The term $1 / \sqrt{2 \omega}$ is treated as a slowly-varying factor, 
which allows us to approximate it by a constant 
in the frequency domain where $f_{\omega_0}(\omega)$ has support.
By definition, 
$\Psi_{\omega_0}(u)$ describes a null wave 
(hence a function of $u$ only) 
centered around $u = 0$. 
Therefore, 
$\Psi_{\omega_0} (u - u_0)$ is an outgoing wave packet 
localized around $u = u_0$ 
with a width of $\Delta u \sim 1/\Delta\omega$ in the $u$-space.
Substituting the solution~\eqref{eq:wavesoln} 
into eq.~\eqref{vaidya_packet} explicitly, 
we obtain
\begin{align}
\widetilde\Psi_{(\omega_0, u_0)} (v, p) 
&= 
\int_0^{\infty} d \omega \, f_{\omega_0}(\omega) \, e^{i \omega u_0} \, 
\mathcal{N}_{\omega} \, 
\frac{e^{-i \omega v}}{p - i 0^+} 
\left[ \frac{ a \left( p - i 0^+ \right)}{\sqrt{1 + \ell^2 p^2}} \right]^{-2 i a \omega} \nn \\
&\approx 
\sqrt{2\omega_0} \, 
\mathcal{N}_{\omega_0} \, \frac{e^{-2 \pi a \omega_0 \Theta (-p)}}{p - i 0^+} \, 
\Psi_{\omega_0} \bigr( v + 2a F(p) - u_0 \bigr)
\, ,
\label{eq:vaidya_packet}
\end{align}
where the slowly-varying factors have again 
been treated as constants and pulled out of the $\omega$-integral. 

We can associate to the Hawking particle 
corresponding to the wave packet~\eqref{vaidya_packet} 
an annihilation operator $b_{\Psi}$ 
defined as the relativistic inner product 
between $\widetilde{\Psi}_{(\omega_0, u_0)}$ 
and the field operator $\tilde{\phi}$, i.e.
\be 
b_{\Psi} \equiv
\bigl\langle \widetilde\Psi_{(\omega_0, u_0)} 
\bigl| \, p \, \bigr|
\tilde{\phi} \bigr\rangle \, .
\ee 
The number expectation value of Hawking particles 
with the wave function $\widetilde\Psi_{(\omega_0, u_0)}$ 
in the Minkowski vacuum $|0\rangle$~\eqref{vacuum}
is then determined by how the operator $b_{\Psi}$ is decomposed 
in terms of the creation and annihilation operators 
$\{ a_p, a_p^{\dagger} \}$ inside the shell.
According to eq.~\eqref{eq:decomp_in0},
this is equivalent to a decomposition into positive and negative-$p$ components.
The calculation of Hawking radiation thus amounts to 
tracing the wave packet $\widetilde\Psi_{(\omega_0, u_0)}$ 
back in time to the collapsing shell at $v = 0$,
where the field $\tilde{\phi}$
can be matched with the mode expansion~\eqref{eq:decomp_in0} inside the shell.
After that, we evaluate the norm of the 
negative-momentum components present in the wave packet.
More explicitly, using the definition in eq.~\eqref{eq:inp}, 
we have
\begin{align}
b_{\Psi} 
&=
\bigl\langle \widetilde\Psi_{(\omega_0, u_0)} (v = 0) 
\bigl| \, p \, \bigr|
\tilde{\phi} (v = 0) \bigr\rangle \nn \\
&= 
\int_0^{\infty} dp \, \sqrt{\frac{p}{2 \left( 1 + \ell^2 p^2 \right)}} 
\left[
\widetilde\Psi_{(\omega_0, u_0)}^* (0, p) \, a_p 
- \widetilde\Psi_{(\omega_0, u_0)}^* (0, -p) \, a_p^{\dagger} 
\right] \, ,
\end{align}
thus the vacuum expectation value of the number of Hawking particles is
\begin{align}
\ev{b_{\Psi}^{\dagger} b_{\Psi}}{0} 
&= 
\int_0^{\infty} dp \, \frac{p}{2(1 + \ell^2 p^2)} \, 
\abs\big{\widetilde\Psi_{(\omega_0, u_0)} (0, -p)}^2
\nn \\
&= \frac{1}{2} \, 
\Bigl\langle \Theta (p) \, \widetilde\Psi_{(\omega_0, u_0)} (0, -p) 
\Bigl| \, p \, \Bigr|
\Theta (p) \, \widetilde\Psi_{(\omega_0, u_0)} (0, -p) \Bigr\rangle \, ,
\label{bb-1}
\end{align}
which is indeed determined by the norm of the negative-momentum components 
in the wave packet $\widetilde\Psi_{(\omega_0, u_0)}$.

To evaluate eq.~\eqref{bb-1},
we first derive the identity
\begin{align}
&\bigl\langle \Theta(p) \, \tilde{\phi}_{\omega} (0, -p) 
\bigl| \, p \, \bigr|
\Theta(p) \, \tilde{\phi}_{\omega'} (0, -p) \bigr\rangle \nn \\
=& \int_0^{\infty} \frac{dp}{1 + \ell^2 p^2} \, \tilde{\phi}_{\omega}^* (0, -p) \, p \, \tilde{\phi}_{\omega'} (0, -p) 
\nn \\
=& \ \mathcal{N}_{\omega}^* \, \mathcal{N}_{\omega'} \, 
e^{- 2 \pi a (\omega \, + \, \omega')} 
\int_0^{\infty} \frac{dp}{p \left( 1 + \ell^2 p^2 \right)} \, 
\exp \left[ 2 i a (\omega - \omega') \log \left( \frac{a p}{\sqrt{1 + \ell^2 p^2}} \right) \right] 
\nn \\
=& \ \mathcal{N}_{\omega}^* \, \mathcal{N}_{\omega'} \, e^{- 2 \pi a (\omega \, + \, \omega')} 
\int_{-\infty}^{\log(a / \ell)} dF \, e^{2 i a (\omega - \omega') F} 
\label{0phiphi0}
\end{align}
by plugging in the mode solutions~\eqref{eq:wavesoln} 
and making a change of variables from $p$ to $F(p)$ 
as introduced in eq.~\eqref{F-def}. 
We observe that the influence of the GUP emerges in the form of an upper bound $\log(a / \ell)$ on the $F$-integration. 
Finally, 
by substituting eq.~\eqref{0phiphi0} into eq.~\eqref{bb-1} 
and making use of eq.~\eqref{N2},
we find the number of Hawking particles 
with the wave function $\widetilde\Psi_{(\omega_0, u_0)}$
in Hawking radiation to be
\begin{align}
\ev{b_{\Psi}^{\dagger} b_{\Psi}}{0} 
&\approx
\frac{a}{2} \, \frac{e^{-4 \pi a \omega_0}}{1 - e^{-4 \pi a \omega_0}} 
\int d\omega \, f_{\omega_0}^* (\omega) \int d \omega' \, f_{\omega_0}(\omega') \, 
e^{-i (\omega - \omega') u_0} \int_{-\infty}^{\log(a / \ell)} dF \, e^{2 i a (\omega - \omega') F} 
\nn \\
&= \frac{1}{4} \, \frac{1}{e^{4 \pi a \omega_0} - 1} 
\int d\omega \, f_{\omega_0}^* (\omega) \int d \omega' \, f_{\omega_0}(\omega') 
\int_{-\infty}^{2 a \log(a / \ell) - u_0} du \, e^{i (\omega - \omega') u} 
\nn \\
&= \frac{1}{2} \, \frac{\omega_0}{e^{4 \pi a \omega_0} - 1} 
\times 
\int_{-\infty}^{2 a \log(a / \ell) - u_0} du \, 
\bigl| \Psi_{\omega_0}(u) \bigr|^2 \, .
\label{VEV-n-1}
\end{align}
It is evident from the Planck distribution factor in the spectrum 
that the Hawking temperature $T_H = 1 / 4 \pi a$ is robust.
It is closely related to the analytic continuation 
of the term $(p - i 0^+)^{-2 i a \omega}$ 
in the expression~\eqref{eq:wavesoln} for the Hawking modes across $p = 0$, 
highlighting its significance as an infrared property.

Given that the wave function $\Psi_{\omega_0}(u)$ 
defined in eq.~\eqref{Psi=Phi-1} 
is centered around $u = 0$ with a width of $\Delta u$,
the integral in eq.~\eqref{VEV-n-1} is highly suppressed when
\be
u_0 - 2a \log(a/\ell) \gtrsim \Delta u \, .
\ee
For a large black hole,
the scrambling time $2a \log(a/\ell)$ is much larger than $\Delta u$,
which is typically of $\mathcal{O}(a)$.\footnote{
$\Delta u$ should be much larger than $a$ so that
the resolution of frequency $\Delta \omega$ is high enough 
to determine the Hawking temperature.
But in the sense of the $\ell/a$-expansion,
we have $\Delta u \sim \mathcal{O}(a)$,
as $\Delta u$ does not scale with $\ell$.
}.
Hence, eq.~\eqref{VEV-n-1} indicates that 
the probability of detecting Hawking particles 
centered around the retarded time $u_0$ diminishes with increasing $u_0$. 
This probability eventually tends to zero when
\be
u_0 \gtrsim 2 a \log(a / \ell) \, ,
\label{u0-scr}
\ee
signifying that 
Hawking radiation is turned off at around the scrambling time.
Although we have essentially followed 
the formulation presented in Ref.~\cite{Brout:1998ei},
an explicit calculation of the time-dependent amplitude of Hawking radiation 
has led to a different conclusion, 
namely, that the GUP eventually shuts down Hawking radiation.

The origin of the termination of Hawking radiation is 
the non-conservation of the inner product 
$\langle \tilde{\phi}_{\omega}(v) | \, p \, | \tilde{\phi}_{\omega'}(v) \rangle$
for the mode solutions~\eqref{eq:wavesoln},
which in fact do not satisfy the boundary condition~\eqref{BC-p}.
This was already pointed out in Ref.~\cite{Brout:1998ei},
which also provided a physical interpretation 
of the non-conservation of the inner product as follows.
In string theory,
when the energy and momentum are trans-Planckian,
there is a plethora of massive modes 
that the massless field can transition into, 
and the nonlocality introduced by the GUP implies 
the dissipation of a Hawking particle into 
a ``reservoir'' containing these massive modes 
when traced backward in time~\cite{Brout:1998ei}. 
Here we offer a similar but slightly different interpretation, 
especially regarding the ultimate outcome of Hawking radiation. 
In our view, 
the trans-Planckian modes with $p \gg a / \ell^2$ 
correspond to long strings with spatial extensions $\Delta x \gg a$.
Hence, 
they are essentially evolving in the asymptotically flat region 
rather than the near-horizon region,
and they end up not contributing to Hawking radiation.
We will elaborate more on this in the ensuing discussion. 

In deriving the action~\eqref{eq:action} for the scalar field in the near-horizon region,
we have made the assumption that $x$ is small, 
as indicated in eq.~\eqref{x<<a}. 
This allowed us to perform a Taylor expansion of the metric 
up to first order in $x / a$ around $x = 0$. 
Subsequently, 
we replaced $x$ with $\hat{x} = i(1 + \ell^2 p^2) \del_p$ 
to incorporate the effects of the GUP.
Strictly speaking, 
this approach is not entirely self-consistent, 
since $\hat{x}$ becomes large as $p$ increases 
due to the presence of the factor $(1 + \ell^2 p^2)$. 
On the other hand, 
the momentum $p$ experiences a significant blueshift only if $x$ is small.
In Appendix~\ref{app:nearhorizon}, 
we address this concern by repeating the calculation 
using the \emph{exact} ingoing Vaidya metric~\eqref{ingoing-Vaidya-metric} 
without relying on the assumption that $x$ is small. 
Remarkably, 
our analysis in the appendix leads to the same conclusion, 
affirming the validity of our findings.

Let us now discuss the physical picture 
behind the termination of Hawking radiation.
Hawking radiation stems from the distinction 
between the notion of particles and vacua defined for 
freely falling observers in the near-horizon region ($x \ll a$)
and distant observers in the asymptotically flat region ($x \gg a$).
In the low-energy effective theory, 
since the Killing vector $\hat{P}_u \equiv i \del_u$ and 
the Kruskal time-derivative operator 
$\hat{P}_U \equiv i \del_U = e^{u / 2a} \hat{P}_u$ 
satisfy the commutator $\comm*{\hat{P}_u}{\hat{P}_U} = i \hat{P}_U / 2a$,
there is an uncertainty relation~\cite{Ho:2021sbi}
\be
\Delta\omega \Delta\Omega 
\geq 
\frac{\langle \omega \rangle}{4a} \, e^{u / 2a} 
\ee
involving their associated frequencies $\omega$ and $\Omega$. 
As the momentum $p = 2 e \Omega \sim \Omega$ is related to 
the Kruskal frequency $\Omega$ by a constant factor of order 1 
on the collapsing shell near the horizon, 
the inequality above implies that a Hawking wave packet 
with dominant frequency 
$\langle \omega \rangle \sim 1 / a \gtrsim \Delta \omega$ 
detected at a later time (larger $u$) by a distant observer 
has a broader distribution of momentum in the past:
\be
\Delta p \gtrsim \langle \omega \rangle \, e^{u / 2a} \sim \langle p \rangle \, ,
\ee
larger than the exponentially blueshifted value $\langle p \rangle$ itself. 
This ensures a mixture of positive and negative-$p$ components 
within a Hawking particle composed of purely positive-$\omega$ modes.
More importantly, 
since the size $\Delta x \sim 1 / \Delta p$ of the wave packet 
shrinks exponentially into the past in the low-energy effective theory, 
it can always be traced back to the near-horizon region 
where the notion of vacuum is different, 
giving rise to Hawking radiation (see eq.~\eqref{bb-1}).

On the other hand,
when the GUP~\eqref{GUP} is introduced,
the uncertainty in position also grows as the momentum increases.
More precisely,
the GUP implies that
\be
\Delta x \gg a
\quad \mbox{when} \quad
\Delta p \gg \frac{a}{\ell^2} \, .
\ee
As a consequence,
for Hawking radiation beyond the scrambling time $u \gtrsim 2 a \log ( a^2 / \ell^2 )$,
the wave packet has a width of $\Delta p \gtrsim \langle p \rangle \sim a / \ell^2$ in the past, 
and thus spreads over a large spatial distance, 
encompassing mainly the asymptotically flat region.
Moreover, 
compared to the exponential blueshift in the low-energy effective theory, 
the characteristic trajectory $v + 2 a F(p) = u_0$~\eqref{eq:chr_traj} 
indicates a substantially more rapid growth in momentum 
during the backward propagation.
In fact, 
Hawking particles that reach the asymptotic region 
after the scrambling time would have originated within the collapsing shell 
with a central momentum $\langle p \rangle \gg a / \ell^2$ far exceeding the Planck scale\,\footnote{
Take the Gaussian profile function 
$f_{\omega_0}(\omega) \propto \exp\left[ - (\omega - \omega_0)^2 / 2 (\Delta \omega)^2 \right]$ 
as an example. 
The width of the wave packet 
$\widetilde{\Psi}_{(\omega_0, u_0)}(v, p)$~\eqref{eq:vaidya_packet} 
in momentum space is then 
$\Delta p \sim \langle p \rangle \left( 1 + \ell^2 \langle p \rangle^2 \right)$. 
Thus, a large central momentum $\langle p \rangle \gg a / \ell^2$ 
implies a much wider spectrum $\Delta p \gg \langle p \rangle$ 
in comparison with the low-energy effective theory.
}. 
Therefore, 
their corresponding wave packets would have spanned an extent 
$\Delta x \gg \mathcal{O}(a)$ far beyond the size of the black hole in the past. 
For these modes, 
the notion of vacuum naturally aligns with 
the vacuum for distant observers, 
rather than the $a_p$-vacuum~\eqref{vacuum} inside the shell. 
This situation remains essentially the same as these highly trans-Planckian, 
\emph{large-scale} modes traverse through the black hole spacetime. 
Hence, these modes are not expected to contribute to particle creation, 
and this is the underlying reason why Hawking radiation 
comes to a stop past the scrambling time. 

\section{GUP in Freely Falling Frames}
\label{sec:ff}

As the implementation of the GUP via eq.~\eqref{hat_x_hat_p}
breaks local Lorentz symmetry,
it depends on the choice of the coordinate system.
In this section, 
we extend our discussions to a class of freely falling frames. 
We find that,
from the viewpoint of a stationary observer at a fixed distance from the horizon,
the conclusion remains the same that
Hawking radiation is turned off when
\be
\Delta t \gtrsim 2a \log(a/\ell) \, ,
\ee
where $\Delta t$ is the duration in Schwarzschild time
starting from the moment when the collapsing matter is at a distance of $\mathcal{O}(a)$ from the horizon.

A particular freely falling frame was considered in previous works 
such as Refs.~\cite{Unruh:1994je, Corley:1996ar} 
for the study of Hawking radiation
with modified dispersion relations.
It is given by the Schwarzschild metric~\eqref{Schwarzschild}
in the Painlevé-Gullstrand coordinates $(\tau, r)$:
\begin{equation}
ds^2 = -d\tau^2 + \left( dx + \sqrt{\frac{a}{r}} \, d\tau \right)^2 \, ,
\label{FFF-1}
\end{equation}
where
\begin{align}
    \tau &\equiv
    t +2 \sqrt{a r} - a \log \left| \frac{\sqrt{r}+\sqrt{a}}{\sqrt{r}-\sqrt{a}} \right| \, ,
    \label{tau-def}
    \\
    x &\equiv r - a \, .
    \label{x-def}
\end{align}
The coordinate $\tau$ agrees with the proper time of a free falling observer
following the trajectory $dx/d\tau = - \sqrt{a/r}$.

We generalize this freely falling frame as follows.
Consider now a generic freely falling observer
who travels radially inward along a timelike geodesic in the Schwarzschild spacetime. 
In terms of the Schwarzschild coordinates $(t, r)$, 
the geodesic equations are 
\be 
\label{eq:geodesic}
\frac{dt}{d\tau} = \frac{E}{1 - a / r} \, , \qquad \left( \frac{dr}{d\tau} \right)^2 = E^2 - 1 + \frac{a}{r} \, ,
\ee 
where $\tau$ is the proper time along the trajectory, 
and $E$ is the conserved energy per unit mass.
For a geodesic that originates from the infinite past ($t \rightarrow - \infty, r \rightarrow \infty$)
with an initial speed given by
\be 
\lim_{t\rightarrow-\infty}\left(\frac{dr}{d t}\right)^2 = 1 - \gamma \, , 
\qquad
 \text{where} \ \gamma \in (0, 1] \, ,
\ee 
we have $E = \gamma^{-1/2}$.
This gives a family of freely falling trajectories 
$\bigl( t(\tau; \gamma), r(\tau; \gamma) \bigr)$ parametrized by $\gamma$. 
The case where $\gamma = 1$ coincides with the special case~\eqref{FFF-1} mentioned above,
which corresponds to an observer initially at rest at past infinity. 
On the other hand, 
the limit $\gamma \to 0$
is closely related to the coordinate system considered in the previous section.  

From the geodesic equations~\eqref{eq:geodesic}, one can introduce the proper time coordinate $\tau(t, r)$ as 
\be 
\tau (t , r) = \gamma^{-1/2} \left[ t + h_{\gamma}(r) \right] \, ,
\label{tau=t}
\ee
where 
\begin{align}
h_{\gamma}(r) &\equiv \int^r dr' \, \frac{\sqrt{1 - \gamma (1 - a / r')}}{1 - a / r'}
\nn \\
\begin{split}
&= \sqrt{1 - \gamma ( 1 - a/r )} \, r 
+ \frac{(2 - \gamma)}{2 \sqrt{1 - \gamma}} \, a \log \left\lvert \frac{\sqrt{1 - \gamma} + \sqrt{1 - \gamma(1 - a / r)}}{\sqrt{1 - \gamma} - \sqrt{1 - \gamma(1 - a / r)}} \right\rvert
\\
&\quad - a \log \left\lvert \frac{1 + \sqrt{1 - \gamma(1 - a / r)}}{1 - \sqrt{1 - \gamma(1 - a / r)}} \right\rvert
\end{split}
\end{align}
up to an irrelevant additive constant. 
Subsequently, eq.~\eqref{tau=t} can be used to define 
a class of free-fall coordinates $(\tau, r)$,
in which the Schwarzschild line element~\eqref{Schwarzschild} 
can be expressed as
\be 
d s^2 = - d \tau^2 + \gamma \bigl( d r - v_{\gamma}(r) \, d \tau \bigr)^2 \, ,
\label{ds2-V}
\ee 
where 
\be 
\label{eq:ff_vel}
v_{\gamma}(r) = - \gamma^{-1/2} \sqrt{1 - \gamma \left( 1 - \frac{a}{r} \right)} \, .
\ee 
In the near-horizon region where $0 < x \ll a$,
we have
\be
d\tau \approx \gamma^{-1/2} \left( dt + dr_{\ast} \right) = \gamma^{-1/2} dv \, .
\label{tau=v}
\ee
Therefore,
through a constant shift of the coordinate $\tau$,
we can approximately identify $\tau$ with $\gamma^{-1/2} v$ in the near-horizon region.

In this coordinate system, 
the low-energy free field theory for a massless scalar has the action
\be 
\label{eq:ff_action}
S_0 = \frac{\gamma^{1/2}}{2} \int d \tau \int d r \left\{
\left[ \left( \del_{\tau} + v_{\gamma}(r) \, \del_r \right) \phi \right]^2 - \gamma^{-1} \left( \del_r \, \phi \right)^2 
\right\} \, .
\ee 
In the near-horizon region
$\abs{x} = \abs{r - a} \ll a$,
eq.~\eqref{eq:ff_vel} reduces to 
$v_{\gamma}(x) \approx - \gamma^{-1/2} + \gamma^{1/2} \, x / 2a$,
so the action~\eqref{eq:ff_action} can be simplified as
\be 
\label{eq:ff_action2}
S_0 = \frac{\gamma^{1/2}}{2} \int d \tau \int d x \left\{ \left[ 
\left( \del_{\tau} - \gamma^{-1/2} \del_x + \gamma^{1/2} \, \frac{x}{2a} \, \del_x \right) \phi \right]^2 
- \gamma^{-1} ( \del_x \, \phi )^2 
\right\} \, .
\ee 
To introduce the GUP into the theory, 
we follow a similar procedure as the previous section
by adopting the representation $\hat{x} = i \left( 1 + \ell^2 p^2 \right) \partial_p$ 
in the momentum space and rewriting the action~\eqref{eq:ff_action2} as
\be 
\label{eq:ff_action_GUP}
S_{\text{GUP}} = 
\gamma^{1/2} \int d \tau \int_0^{\infty} \frac{dp}{1 + \ell^2 p^2} 
\left\{ 
\left| 
\left( \del_{\tau} - i \gamma^{-1/2} p - \gamma^{1/2} \, \frac{1 + \ell^2 p^2}{2 a} \, \del_p \, p \right) 
\tilde{\phi} 
\right|^2 
- \gamma^{-1} \, p^2 \left|\tilde{\phi}\right|^2
\right\} \, ,
\ee 
where the measure factor $1 / \left(1 + \ell^2 p^2\right)$ is again required in the $p$-space
for $\hat{x}$ to be Hermitian. 
The corresponding equation of motion reads
\be 
\label{eq:ff_eom}
\left(\del_{\tau} - 2i \gamma^{-1/2} p - \gamma^{1/2} \, \frac{1 + \ell^2 p^2}{2a} \, p \, \del_p \right)
\left(\del_{\tau} - \gamma^{1/2} \, \frac{1 + \ell^2 p^2}{2a} \, \del_p \, p \right)
\tilde{\phi}(\tau, p) = 0 \, .
\ee 
In particular, the wave equation for the outgoing modes $\tilde{\phi}_{\text{out}}(\tau, p)$ is
\be
\left(\del_{\tau} - \gamma^{1/2} \, \frac{1 + \ell^2 p^2}{2a} \, \del_p \, p \right)
\tilde{\phi}_{\text{out}}(\tau, p) = 0 \, .
\label{outgoing-wave-eq-FFF}
\ee
Notice that if we identify $\del_{\tau}$ with $\gamma^{1/2} \del_v$
using the relation~\eqref{tau=v} in the near-horizon region,
eq.~\eqref{outgoing-wave-eq-FFF} coincides precisely with
the outgoing wave equation~\eqref{outgoing-wave-eq}
in the previous section
despite the difference in the full wave equations 
between the two scenarios.
Hence,
the general solution to eq.~\eqref{outgoing-wave-eq-FFF} has the same structure as eq.~\eqref{outgoing-genera-sol}:
\be
\tilde{\phi}_{\text{out}}(\tau, p) = p^{-1} \, \Psi \bigl( \gamma^{1/2} \tau + 2aF(p) \bigr) \, ,
\label{general-sol}
\ee
where $F(p)$ is again defined by eq.~\eqref{F-def}.

In the previous section, 
we recognized based on eqs.~\eqref{eq:decomp_in0} and~\eqref{vacuum} 
that positive/negative-momentum modes match with
positive/negative-frequency modes 
in the Minkowski space inside the collapsing shell 
in the sense that they share the same notion of vacuum. 
This correspondence persists in freely falling frames,
and it can be established by noting that 
\be
\del_U \approx 
\frac{1}{2} \, e^{\gamma^{1/2} \tau / 2a}
\left(\del_{\tau} + v_{\gamma}(x) \, \del_x - \gamma^{-1/2} \del_x \right)
\ee
in the near-horizon region,
where $U = - 2a e^{- u / 2a}$ is the Kruskal retarded time.
Acting $\del_U$ on the general solution~\eqref{general-sol} for outgoing modes gives
\be
\Omega \approx \gamma^{-1/2} \, e^{\gamma^{1/2} \tau / 2a} \, p \, ,
\label{Om-p}
\ee
with $\Omega$ denoting the eigenvalue of $i \del_U$.
The $\tau$-dependent factor in the equation above is positive-definite 
and corresponds to the exponential redshift $e^{- \gamma^{1/2} \tau / 2a}$ 
of the momentum $p$ in the near-horizon region. 
Thus according to eq.~\eqref{Om-p},
positive-$\Omega$ modes can be identified with 
positive-$p$ modes in the near-horizon region, 
and vice versa. 

Given that the wave equation~\eqref{outgoing-wave-eq-FFF} for outgoing modes
can be made identical to that in the previous section 
through a (positive) rescaling $\tau \approx \gamma^{-1/2} v$~\eqref{tau=v} of time, 
and considering that the decomposition into positive/negative-$\Omega$ modes 
and into positive/negative-$p$ modes 
are equivalent within the family of freely falling frames 
just as in the previous section, 
we anticipate that the implementation of the GUP 
with respect to the freely falling coordinates 
would yield results for Hawking radiation 
identical to those discussed in the prior section 
for all values of $\gamma \in (0, 1]$. 
Nevertheless, 
for the sake of thoroughness, 
we proceed to replicate the same sequence of steps 
presented in the previous section.

The single-frequency solutions can be derived from eq.~\eqref{general-sol} as
\be 
\label{eq:ff_sol}
\tilde{\phi}_{\omega} (\tau, p) = 
\mathcal{N}_{\omega} \, 
\frac{e^{-i \omega \tau}}{p - i 0^+} 
\left[ \frac{ a \left( p - i 0^+ \right)}{\sqrt{1 + \ell^2 p^2}} \right]^{-2 i a \gamma^{-1/2} \omega} \, .
\ee 
According to eq.~\eqref{tau=t},
for a stationary observer at fixed $r$,
the frequency $\omega$ defined with respect to $\tau$
corresponds to a frequency $\gamma^{-1/2} \omega$ 
with respect to the Eddington advanced time $v$
(as well as the Schwarzschild time $t$).
Hence,
in order for $\phi_{\omega}(\tau, x)$ to match
the outgoing mode 
$e^{- i \gamma^{-1/2} \omega u} / \sqrt{2 \gamma^{-1/2} \omega}$ at large $\tau$,
it can be verified that the normalization constant $\mathcal{N}_{\omega}$ 
is again given by eq.~\eqref{N-def}, 
but with $\omega$ replaced by $\gamma^{-1/2}\omega$.
Through this rescaling of the frequency,
eq.~\eqref{eq:ff_sol} is in agreement with eq.~\eqref{eq:wavesoln}. 

To define the vacuum state for freely falling observers,
consider the local patch of spacetime near the worldline of a freely falling observer 
comoving with the collapsing matter.
In the near-horizon region, 
we can associate to the local patch 
a set of coordinates $(T, X)$ defined by
(see eq.~\eqref{ds2-V})
\begin{align}
d T &= d \tau \, , 
\label{ddT} \\
d X &\approx \gamma^{1/2} \left[ d x - v_{\gamma}(x) \, d \tau \right]
\approx \gamma^{1/2} d x + d\tau 
\, .
\label{ddX}
\end{align}
The freely falling observer close to the horizon 
would then adopt the free-field action~\eqref{eq:ff_action_GUP} in the local patch expressed as
\be 
S_{\text{GUP}} = \gamma^{1/2} \int dT \int_0^{\infty} \frac{dp}{1 + \ell^2 p^2} 
\left(
\left| \del_{\, T} \, \tilde{\phi} \, \right|^2 
- \gamma^{-1} \, p^2 \left| \tilde{\phi}\right|^2 \, \right) \, ,
\label{S-flat}
\ee 
and the general solution to the field equation for this action is
\be 
\label{eq:ff_decomp}
\tilde{\phi}(T, p) = 
\sqrt{\frac{1 + \ell^2 p^2}{2 \abs{p}}} \, e^{-i \gamma^{-1/2} p \, T}
\left[ a_p \, \Theta(p) 
+ a_{-p}^{\dagger} \, \Theta(-p) 
\right] 
+ \mbox{ingoing modes} \, ,
\ee 
where $a_p$ and $a_p^{\dagger}$ are the creation and annihilation operators 
associated with the outgoing modes. 

\begin{figure}[t]
\centering
\includegraphics[scale=0.9]{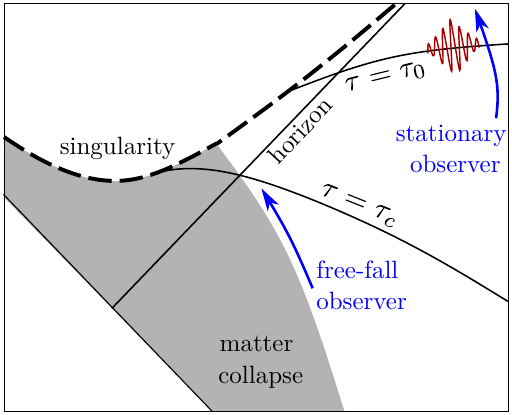}
\caption{\label{fig:freefall}
The Kruskal diagram of a Schwarzschild black hole formed from collapse, 
foliated by hypersurfaces of constant $\tau$. 
The time at which the horizon emerges is the crossing time denoted by $\tau_c$.
A Hawking wave packet detected by a stationary observer around $\tau = \tau_0$ is traced back in time and decomposed into positive and negative-frequency components with respect to a freely falling observer comoving with the surface of the collapsing matter.
}
\end{figure}

To compute Hawking radiation,
we trace the wave packet of a Hawking particle
backward in time to the near-horizon region
and calculate the norm of its negative-frequency components
with respect to a freely falling observer. 
A generic wave packet for a Hawking particle centered at $\tau = \tau_0$ 
has the form
\be 
\label{eq:ff_packet}
\widetilde{\Psi}_{(\omega_0, \tau_0)}(\tau, p) = 
\int_0^{\infty} d \omega \, f_{\omega_0}(\omega) \, e^{i \omega \tau_0} \, 
\tilde{\phi}_{\omega}(\tau, p) \, ,
\ee 
which is analogous to eq.~\eqref{vaidya_packet},
but defined in a different reference frame.
At large $\tau$,
the wave packet conforms to an ordinary null packet 
of the low-energy effective theory, 
and similar to eq.~\eqref{Psi=Phi-1} in the previous section, 
its wave function in the position space is approximately a function of $u$ only:
\be
\Psi_{\omega_0}(u - u_0) =
\frac{1}{\sqrt{2 \gamma^{-1/2} \omega_0}}
\int_0^{\infty} d \omega \, f_{\omega_0}(\omega) \, 
e^{- i \gamma^{-1/2} \omega (u - u_0)} \, ,
\label{Psi=Phi-2}
\ee
where $u_0 \equiv \gamma^{1/2} \tau_0$. 

The wave packet~\eqref{eq:ff_packet} is traced back in time to be matched with 
the free-fall mode expansion~\eqref{eq:ff_decomp} 
at around the crossing time $\tau_c$ 
(denoted as $T_c$ in the coordinate patch of a freely falling observer)
when the collapsing matter is about to 
pass through the horizon (see figure~\ref{fig:freefall}). 
With $b_{\Psi}$ representing the annihilation operator 
associated with the Hawking particle described by this wave packet, 
we obtain from the relativistic inner product (see eq.~\eqref{eq:inp})
\begin{align}
b_{\Psi} 
&= \bigl\langle \widetilde\Psi_{(\omega_0, \tau_0)} (\tau_c) 
\bigl| \, p \, \bigr|
\tilde{\phi} (T_c) \bigr\rangle \nn \\
&= \int_0^{\infty} dp \, \sqrt{\frac{p}{2 \left( 1 + \ell^2 p^2 \right)}} 
\left[
\widetilde{\Psi}_{(\omega_0, \tau_0)}^*(\tau_c \, , p)  \,
e^{-i \gamma^{-1/2} \, p \, T_c} \, a_p 
- \widetilde{\Psi}_{(\omega_0, \tau_0)}^*(\tau_c \, , - p) \,
e^{i \gamma^{-1/2} \, p \, T_c} \, a_p^{\dagger} 
\right] \, .
\end{align}
Assuming the free-fall vacuum state $\ket{0}$ (Unruh vacuum) 
annihilated by $a_p$ for all positive $p$, 
we determine the number of Hawking quanta by following 
the same steps outlined in the previous section, 
leading to
\begin{align}
\ev{b_{\Psi}^{\dagger} b_{\Psi}}{0} 
&= 
\frac{1}{2} \int_0^{\infty} dp \, \frac{p}{1 + \ell^2 p^2} \, 
\abs\big{\widetilde{\Psi}_{(\omega_0, \tau_0)} (\tau_c, -p)}^2 
\nn \\
&\approx 
\frac{a / 2}{e^{4 \pi a \gamma^{-1/2} \omega_0} - 1} \int d\omega \, d \omega' \, 
f_{\omega_0}^* (\omega) \, f_{\omega_0}(\omega') \, 
e^{-i (\omega - \omega') (\tau_0 - \tau_c)} 
\int_{-\infty}^{\log(a / \ell)} dF \, e^{2 i a \gamma^{-1/2} (\omega - \omega') F} 
\nn \\
&= 
\frac{1}{2} \, \frac{\gamma^{-1/2} \omega_0}{e^{4 \pi a \gamma^{-1/2} \omega_0} - 1} 
\times 
\int_{-\infty}^{2 a \gamma^{-1/2} \log(a / \ell) - \Delta \tau} d\tau \, \gamma^{1/2} \,
\bigl| \Psi_{\omega_0}(\gamma^{1/2} \tau) \bigr|^2
\nn \\
&= 
\frac{1}{2} \, \frac{\gamma^{-1/2} \omega_0}{e^{4 \pi a \gamma^{-1/2} \omega_0} - 1} 
\times 
\int_{-\infty}^{2 a \log(a / \ell) - \Delta t} dt \,
\bigl| \Psi_{\omega_0}(t) \bigr|^2 \, ,
\label{eq:ff_VEV}
\end{align}
where $\Delta \tau \equiv \tau_0 - \tau_c$\,,
$\Delta t \equiv \gamma^{1/2} \Delta \tau$ is the difference in the Schwarzschild time $t$,
and $\gamma^{- 1/2}\omega_0$ is the central frequency 
of the Hawking particle defined with respect to $t$.
As it turns out, 
the Hawking radiation is turned off when 
$\Delta \tau \gtrsim 2 a \gamma^{-1/2} \log(a / \ell)$.
For a distant observer situated at a fixed radius $r$,
this shutdown time scale can be expressed as
\be
\Delta t \gtrsim 2 a \log(a / \ell) \, .
\label{Deltat}
\ee 
This is in agreement with the finding~\eqref{u0-scr} presented in the previous section, 
as $\Delta t$ coincides with $\Delta u$ at a fixed $r$.
Strikingly, 
despite the fact that the implementation of the GUP 
relies on the choice of coordinates, 
the final outcome~\eqref{eq:ff_VEV}, 
from the viewpoint of a stationary observer, 
is independent of the parameter $\gamma$,
which distinguishes different freely falling frames.

\section{Conclusion and Discussion}
\label{sec:conclusion}

In this work,
we studied the effect of the generalized uncertainty principle (GUP)
on Hawking radiation 
in both the Eddington-Finkelstein coordinates 
and a class of freely falling frames
adapted to geodesic observers 
with different initial velocities at past infinity.
Our approach involved deforming 
the radial coordinate of the black hole background 
in a way that realizes the minimal length uncertainty relation~\eqref{GUP}. 
We came to the conclusion that due to the GUP,
Hawking radiation is eventually terminated 
at around the scrambling time $\Delta t \sim 2a \log(a / \ell)$, 
as measured by a distant observer who is stationary with respect to the black hole.
Remarkably, 
this outcome turns out to be independent of 
the specific choice of the coordinate system in which the GUP is implemented.

The primary reason that led to the eventual decrease in Hawking radiation 
at late times is the following.
Notice that the effect of the GUP essentially boils down to 
a substitution of the momentum $p$ with an ``effective momentum'' $q(p)$ 
in the evaluation of the number expectation value of Hawking particles 
(see, for example, eq.~\eqref{0phiphi0}). 
This change of variables is encapsulated by the relation
\be
p \rightarrow q(p) \equiv \frac{p}{\sqrt{1+\ell^2 p^2}} \, ,
\label{q-def}
\ee
where the unbounded interval of $p$ is mapped to a finite range of $q$, 
with a cutoff value given by $q(\infty) = \ell^{-1}$. 
Consequently, 
the situation is equivalent to having a UV cutoff $\ell^{-1}$ 
imposed on the frequency spectrum, 
which causes a significant reduction in the amplitude of Hawking radiation 
beyond the scrambling time, a result consistent with the conclusions drawn from earlier works~\cite{Ho:2021sbi, Ho:2022gpg}.

Nevertheless, 
we emphasize that the physical interpretation 
underlying the scenario above with the GUP 
greatly differs from a mere truncation of the effective field theory via a UV cutoff. 
In our implementation of the GUP, 
we did not exclude the presence of trans-Planckian modes.
Instead, 
these modes are simply not expected to play a role in the Hawking process, 
as elaborated upon in the end of Sec.~\ref{sec:vaidya}.
The reason behind this lies in the fact that 
the trans-Planckian modes exhibit a large uncertainty 
$\Delta p \gg a  / \ell^2$ in momentum, 
which in turn gives rise to a large uncertainty 
$\Delta x \gtrsim \ell^2 p \gg a$ in position implied by the GUP~\eqref{GUP}. 
Given that the length scales associated with these modes are significantly larger 
than the size of the black hole, 
their evolution is more suitably described 
in the asymptotically flat region far from the black hole, 
and thus do not lead to particle creation.
As a result, 
the late-time Hawking radiation, 
which would have originated from trans-Planckian fluctuations 
in the vacuum near the horizon,
would no longer exist once the effects of the GUP are accounted for.

Since Hawking radiation lasts for approximately the scrambling time, 
only a fraction of the energy of order $\mathcal{O}\bigl( (\ell^2 / a^2) \log(a / \ell) \bigr)$ relative to the original black hole mass will be evaporated away.
Therefore, 
the information loss problem, 
which typically arises around the Page time 
$\mathcal{O}(a^3 / \ell^2)$~\cite{Page:1993wv}, 
is absent here,
and the firewall~\cite{Almheiri:2012rt} does not emerge.
The emitted radiation is thermal and carries no information.
In this sense, 
incorporating the GUP into the radiation field 
effectively results in a \emph{macroscopic} black hole remnant 
that is essentially classical.
This scenario also avoids the issues associated with 
Planck-size remnants as the end state of evaporation 
(see, e.g.,~\cite{Chen:2014jwq, Ong:2018zqn}).
Moreover,
these findings have significant implications 
in the context of primordial black holes in the early universe.
Considering the substantial distinctions 
in both the evaporation time scale 
and the relic masses of primordial black holes under the GUP,
we anticipate a completely different number density 
of black hole remnants produced at the end of inflation.
This would impact the proposal that primordial black hole remnants could
serve as candidates for dark matter~\cite{MacGibbon:1987my, Chen:2002tu, Baumann:2007yr}. 

This study was inspired by a series of investigations that challenged 
the robustness of Hawking radiation against modifications of UV physics.
Notably, it was highlighted in Refs.~\cite{Ho:2020cbf, Ho:2020cvn, Ho:2021sbi, Ho:2022gpg} 
that non-renormalizable higher-derivative interactions, 
while featuring coupling constants suppressed by inverse powers of the Planck mass, 
can induce exponentially large deviations in Hawking radiation 
that lead to a breakdown of the low-energy effective theory around the scrambling time.
Furthermore, 
a more recent work~\cite{Akhmedov:2023gqf} demonstrated that 
Hawking radiation could stop at around the scrambling time, 
at a later time, or persist until the black hole becomes microscopic, 
depending on the UV behavior of modified dispersion relations.
In the present study, 
we focused on the implications of the GUP,
which was originally proposed to capture a certain aspect of string theory.
The outcomes of this work confirmed once again that 
the late-time behavior of the amplitude of Hawking radiation 
is sensitive to UV physics, 
although the Hawking temperature remains robust 
up to perturbative corrections. 
A more comprehensive check is still required 
to ascertain whether this is indeed a prediction of string theory.
On a broader scope, 
it is appealing to further explore how 
the amplitude of Hawking radiation behaves 
under different candidates of quantum gravity.

Much like other theories incorporating a minimum length, 
a theory involving the GUP is inherently nonlocal, 
and thus a large correction to the conventional Hawking radiation
is perhaps not very surprising.
On the other hand,
Hawking radiation in nonlocal theories with a minimum length 
have been examined before in Refs.~\cite{Corley:1997ef, Jacobson:1999ay, Kajuri:2018myh, Boos:2019vcz}, which all reported a robust Hawking radiation.
The reason why the UV-dependence of Hawking radiation 
has been missed in the past can be partially attributed to 
the predominant emphasis on calculating the Hawking temperature, 
a feature indeed resilient to modifications, 
while overlooking the magnitude of the radiation, 
which can undergo substantial changes over time.\footnote{
In addition to the time-dependent nature 
of the amplitude illustrated in this work,
there are other intriguing insights into Hawking radiation 
which the temperature alone does not reveal, 
such as the origin and properties of 
the emission of Hawking pairs examined in Ref.~\cite{Ong:2020hti}.
}
Another reason is that there are subtleties involved 
in the notion of a minimum length.
Future studies aimed at illuminating the connection between 
Hawking radiation and UV theories will be valuable.

\section*{Acknowledgement}

We thank Chi-Ming Chang, Yosuke Imamura, Henry Liao, Nobuyoshi Ohta, Naritaka Oshita, and Yuki Yokokura 
for valuable discussions. 
T.L.C., P.M.H., W.H.S., and C.T.W. are supported in part 
by the Ministry of Science and Technology, R.O.C.
(MOST 110-2112-M-002-016-MY3),
and by National Taiwan University. 
H.K. thanks Prof. Shin-Nan Yang and his family
for their kind support through the Chin-Yu chair professorship.
H.K. is partially supported by Japan Society of Promotion of Science (JSPS),
Grants-in-Aid for Scientific Research (KAKENHI)
Grants No.\ 20K03970 and 18H03708,
by the Ministry of Science and Technology, R.O.C. (MOST 111-2811-M-002-016),
and by National Taiwan University.

\appendix

\section{Beyond the Near-Horizon Approximation}
\label{app:nearhorizon}

In our formulations of GUP-inspired field theory 
in different coordinate systems of a black hole background, 
we employed the near-horizon approximation by 
expanding the spatial dependence around $x = 0$
and retaining terms up to $\mathcal{O}(x / a)$ 
before analyzing the theory in momentum space.
In the ingoing Vaidya coordinates (see Sec.~\ref{sec:vaidya}), 
this yielded the field equation
\be 
\label{eq:app_vaidyaeqn}
\left[ \del_v + \frac{1}{2} \left( 1 - \frac{a}{\hat{x} + a} \right) i p \right] \tilde{\phi}(v, p) \approx \left( \del_v + \frac{\hat{x}}{2a} \, i p \right) \tilde{\phi}(v, p) = 0 \, ,
\ee 
whereas in the freely falling coordinates (see Sec.~\ref{sec:ff}), 
it led to 
\be 
\label{eq:app_ffeqn}
\left[ \del_{\tau} + v_{\gamma}(\hat{x}) \, i p + \gamma^{-1/2} i p \right] \tilde{\phi}(\tau, p) \approx \left( \del_{\tau} + \gamma^{1/2} \, \frac{\hat{x}}{2a} \, i p \right) \tilde{\phi}(\tau, p) = 0 
\ee 
for the outgoing modes.
However, as we implement the representation $\hat{x} = i(1 + \ell^2 p^2) \del_p$, the $\hat{x} / a$-expansion is no longer a valid approximation at large $\abs{p} \gg a / \ell^2$ due to the factor $(1 + \ell^2 p^2)$ introduced by GUP corrections. 
In this appendix, 
we solve the complete field equation 
using the exact Vaidya line element~\eqref{ingoing-Vaidya-metric},
without relying on the $x/a$-expansion. 
We find that our conclusion regarding 
the late-time suppression of Hawking radiation still holds.
A similar scenario is expected to apply 
to the modified field equation~\eqref{eq:app_ffeqn} in the freely falling coordinates.

Let us reframe our analysis in Sec.~\ref{sec:vaidya} 
without restricting to the near-horizon region.
Starting with the conventional action~\eqref{eq:action0} in the low-energy effective theory now expressed as
\be 
S_0 = \frac{1}{2} \int dv \int dr \, \phi(v, r) \, \del_r \, r^{-1} \bigl\{ 2 r \, \del_v + \left[ r - \Theta(v) \, a \right] \del_r \bigr\} \, \phi(v, r) \, ,
\ee 
we arrive at the equivalent form in the momentum space as
\be 
\label{eq:app_action0}
S_0 = \int dv \int_{-\infty}^{\infty} dp \, \tilde{\phi}^*(v, p) \, p \, \hat{r}^{-1} \left\{ i \hat{r} \, \del_v - \left[ \hat{r} - \Theta(v) \, a \right] \frac{p}{2} \right\} \tilde{\phi}(v, p) \, .
\ee 
We introduce the GUP into the radial coordinate $r$ by adopting the representation
\be 
\hat{r} \equiv i f(p) \, \del_p  \, , \qquad \text{where} \quad f(p)  = 1 + \ell^2 p^2 \, .
\ee 
In addition, 
we define the inverse $\hat{r}^{-1}$ as $\hat{r}^{-1} \equiv - i \, \del_p^{-1} f^{-1}(p)$. 
With the incorporation of the GUP, 
the momentum-space action~\eqref{eq:app_action0} becomes 
\be 
S_{\text{GUP}} = 
2 \int dv \int_0^{\infty} \frac{dp}{f(p)} \,  
\tilde{\phi}^*(v, p) \, p \, 
\del_p^{-1} f^{-1}(p) 
\left[ f(p) \, \del_p \left( i \del_v - \frac{p}{2} \right) - \Theta(v) \, \frac{i a}{2} \, p \right]
\tilde{\phi}(v, p) \, ,
\ee 
which leads to the field equation 
\be
\label{eq:app_waveeqn}
\left[ f(p) \, \del_p \left( i \del_v - \frac{p}{2} \right) - \Theta(v) \, \frac{i a}{2} \, p \right]
\tilde{\phi}(v, p) = 0
\ee
for the outgoing modes. 

Outside the shell ($v > 0$), 
the stationary solutions take on the form
\be 
\label{eq:app_soln}
\tilde{\phi}_{\omega}(v, p)
= 
\mathcal{N}_{\omega} \, e^{-i \omega v} \, \exp\left[ -i \, \frac{a \tan^{-1}(\ell p)}{\ell \left( 1 + 4 \ell^2 \omega^2 \right)} \right] \times \frac{1}{p - 2 \omega - i 0^+} \left[ \frac{a \left( p - 2 \omega - i 0^+ \right)}{\sqrt{1 + \ell^2 p^2}} \right]^{- \frac{2 i a \omega}{1 + 4 \ell^2 \omega^2}} \, ,
\ee 
with the constant $\mathcal{N}_{\omega}$ once again determined by 
matching a low-frequency ($\omega \ell \ll 1$) wave packet 
with the standard outgoing wave packet of the low-energy effective theory 
at large $v \gg u_0$, just as in eq.~\eqref{Psi=Phi-1}. 
For low-energy Hawking particles with $\omega \ll \ell^{-1}$, 
this yields the same expression for $\mathcal{N}_{\omega}$ 
as shown in eq.~\eqref{N-def}.

In the flat spacetime inside the shell ($v < 0$), 
the field equation~\eqref{eq:app_waveeqn} reduces to 
\be 
\left( 1 + \ell^2 p^2 \right) \del_p \left( i \del_v - \frac{p}{2} \right) \tilde{\phi}(v < 0, p) = 0 \, ,
\ee 
and thus the outgoing sector of the field has the same mode expansion as in eq.~\eqref{eq:decomp_in0}. 
As a result, 
the rest of the calculation proceeds in a straightforward manner, 
resembling the steps taken in Sec.~\ref{sec:vaidya}. 
In particular, 
in place of eq.~\eqref{0phiphi0},
we now have
\begin{align}
&\bigl\langle \Theta(p) \, \tilde{\phi}_{\omega} (0, -p) 
\bigl| \, p \, \bigr|
\Theta(p) \, \tilde{\phi}_{\omega'} (0, -p) \bigr\rangle \nn \\
=& \ \mathcal{N}_{\omega}^* \, \mathcal{N}_{\omega'} \, 
\exp\left[ - 2 \pi a \left( \frac{\omega}{1 + 4 \ell^2 \omega^2} + \frac{\omega'}{1 + 4 \ell^2 \omega'^{\,2}} \right) \right]
\int_0^{\infty} \frac{dp}{1 + \ell^2 p^2} \, 
\frac{p \, \exp \left[ 2 i a (\omega - \omega') F_{\omega\omega'}(p) \right]}{(p + 2 \omega) (p + 2 \omega')} \nn \\
=& \ \mathcal{N}_{\omega}^* \, \mathcal{N}_{\omega'} \, 
\exp\left[ - 2 \pi a \left( \frac{\omega}{1 + 4 \ell^2 \omega^2} + \frac{\omega'}{1 + 4 \ell^2 \omega'^{\,2}} \right) \right] 
\int_{F_{\omega\omega'}(0)}^{F_{\omega\omega'}(\infty)} dF_{\omega\omega'} \, e^{2 i a (\omega - \omega') F_{\omega\omega'}} \, ,
\label{eq:app_inp}
\end{align}
where
\be 
\begin{aligned}
F_{\omega\omega'}(p) \equiv \frac{1}{\omega - \omega'} \, \Biggl\{ &\frac{\omega}{1 + 4 \ell^2 \omega^2} \, \log \left[ \frac{a \left( p + 2 \omega \right)}{\sqrt{1 + \ell^2 p^2}} \right] - \frac{\omega'}{1 + 4 \ell^2 \omega'^{\,2}} \, \log \left[ \frac{a \left( p + 2 \omega' \right)}{\sqrt{1 + \ell^2 p^2}} \right] \\
&+ \frac{2 \ell \left( \omega^2 - \omega'^{\,2} \right) \tan^{-1}\left( \ell p \right)}{\left( 1 + 4 \ell^2 \omega^2 \right) \left( 1 + 4 \ell^2 \omega'^{\,2} \right)} \Biggr\} \, .
\end{aligned}
\ee 
Similar to the change of variables from $p$ to $F(p)$ in eq.~\eqref{0phiphi0}, 
the construction of $F_{\omega\omega'}(p)$ above 
serves the purpose of simplifying the $p$-integral in eq.~\eqref{eq:app_inp}, 
taking advantage of the fact that 
\be 
dF_{\omega\omega'} = \frac{1}{1 + \ell^2 p^2} \, \frac{p}{(p + 2 \omega) (p + 2 \omega')} \, dp \, .
\ee 
For a large black hole, 
the typical frequency of Hawking radiation lies 
in the range $\omega, \omega' \sim 1 / a \ll \ell^{-1}$, 
which implies that the upper and lower bounds 
of the $F_{\omega\omega'}$-integration in eq.~\eqref{eq:app_inp} 
are approximately given by
\be 
F_{\omega\omega'}(\infty) \approx \log(a / \ell) 
\qquad \text{and} \qquad 
F_{\omega\omega'}(0) \approx \frac{\omega \log(2a \omega) - \omega' \log (2 a \omega')}{\omega - \omega'} \sim \mathcal{O}(1) \, ,
\ee 
respectively. 
Subsequently, 
eq.~\eqref{VEV-n-1} is replaced by
\begin{align}
\ev{b_{\Psi}^{\dagger} b_{\Psi}}{0} 
&\approx \frac{a}{2} \, \frac{e^{-4 \pi a \omega_0}}{1 - e^{-4 \pi a \omega_0}} \int d\omega \, f_{\omega_0}^* (\omega) \int d \omega' \, f_{\omega_0}(\omega') \, e^{-i (\omega - \omega') u_0} \int_{F_{\omega\omega'}(0)}^{\log(a / \ell)} dF_{\omega\omega'} \, e^{2 i a (\omega - \omega') F_{\omega\omega'}} 
\nn \\
&= \frac{1}{4} \, \frac{1}{e^{4 \pi a \omega_0} - 1}  \int d\omega \, f_{\omega_0}^* (\omega) \int d \omega' \, f_{\omega_0}(\omega') \int_{2a F_{\omega\omega'}(0) - u_0}^{2 a \log(a / \ell) - u_0} du \, e^{i (\omega - \omega') u} 
\nn \\
&\approx \frac{1}{2} \, \frac{\omega_0}{e^{4 \pi a \omega_0} - 1} 
\times 
\int_{2 a \, \mathcal{O}(1) - u_0}^{2 a \log(a / \ell) - u_0} du \, 
\bigl| \Psi_{\omega_0}(u) \bigr|^2 \, .
\label{eq:VEV}
\end{align}
We observe that the Hawking radiation is gradually turned on 
as $u_0$ exceeds 
$2a \, \mathcal{O}(1) \sim \mathcal{O}(a)$. 
This transient phenomenon of Hawking radiation becomes evident
only when we extend beyond the near-horizon approximation~\cite{Akhmedov:2023gqf}.
Most importantly, 
as the integral in~\eqref{eq:VEV} is highly suppressed when
\be
u_0 \gtrsim 2 a \log(a / \ell) \, ,
\ee
we reach the same conclusion as in Sec.~\ref{sec:vaidya} 
that Hawking radiation is terminated after the scrambling time.

\small

\bibliographystyle{myJHEP}
\bibliography{bibliography}

\end{document}